%% file: main.tex
\documentclass[reprint, letterpaper, nobibnotes, aps,superscriptaddress,prbt]{revtex4-1}

\usepackage{graphicx}
\usepackage{amsmath}
\usepackage{amssymb}
\usepackage{graphicx}
\usepackage{url}
\usepackage{soul}
\usepackage{hyperref}
\usepackage{xcolor}
\usepackage{subfig}

\begin{document}


\title{A Simple Electrostatic Model for the Hard-Sphere Solute Component of Nonpolar Solvation} 

\author{Christopher D. Cooper}
\affiliation{Dept. of Mechanical Engineering and Centro Cient\'ifico Tecnol\'ogico de Valpara\'iso, Universidad T\'ecnica Federico Santa Mar\'{i}a, Valpara\'{i}so, Chile}
\email[]{christopher.cooper@usm.cl}
\author{Jaydeep P. Bardhan}
\affiliation{GlaxoSmithKline, Philadelphia, USA}
\email[]{jaydeep.p.bardhan@gsk.com}

\date{\today}

\begin{abstract}
We propose a new model for estimating the free energy of forming a molecular cavity in a solvent, by assuming this energy is dominated by the electrostatic energy associated with creating the static (interface) potential inside the cavity. The new model approximates the cavity-formation energy as that of a shell capacitor: the inner, solute-shaped conductor is held at the static potential, and the outer conductor (at the first solvation shell) is held at zero potential. Compared to cavity energies computed using free-energy pertubation with explicit-solvent molecular dynamics, the new model exhibits surprising accuracy (Mobley test set, RMSE 0.45 kcal/mol).  Combined with a modified continuum model for solute-solvent van der Waals interactions, the total nonpolar model has RMSE of 0.55 kcal/mol on this test set, which is remarkable because the two terms largely cancel.  The overall nonpolar model has a small number of physically meaningful parameters and compares favorably to other published models of nonpolar solvation. Finally, when the proposed nonpolar model is combined with our solvation-layer interface condition (SLIC) continuum electrostatic model, which includes asymmetric solvation-shell response, we predict solvation free energies with an RMS error of 1.35~kcal/mol relative to experiment, comparable to the RMS error of explicit-solvent FEP (1.26~kcal/mol). Moreover, all parameters in our model have a clear physical meaning, and employing reasonable temperature dependencies yields remarkable correlation with solvation entropies.
\end{abstract}



\maketitle 


\section{Introduction}
\input{intro.tex}

\section{Theory and Methods}
\input{methods.tex}

\section{Results and Discussion}
\input{results.tex}

\section{Summary and Future Work}
\input{conclusion.tex}

\section*{Supporting Information}
The Supporting Information includes a detailed derivation of the capacitor model as well as figures with additional results named in the text, and other plots establishing the accuracy of the numerical calculations.  
Readers interested in exploring the relationship between our computational results and the reference explicit-solvent FEP calculations of Mobley {\it et al.} \cite{MobleyETal2009} may download a Jupyter Notebook and all the data from \url{https://github.com/cdcooper84/nonpolar_solvation}.
The Github repository also contains code to reproduce the results of this paper. 
\section*{Data Availability Statement}
The results of the calculations in this study, and the code to reproduce the results, are openly available in \url{https://github.com/cdcooper84/nonpolar_solvation}. The MD FEP data of Mobley et al. ~\cite{MobleyETal2009} are available upon reasonable request from the authors of that work, and will soon be openly available via DOI.

\begin{acknowledgments}
The authors thank D. Mobley for sharing simulation results, and gratefully acknowledge valuable discussions with M. Knepley, N. Baker, M. Schnieders, P. Ren, D. Rogers, and M. Radhakrishnan.  This work was supported by CONICYT-Chile through FONDECYT Iniciaci\'on N$^\circ$ 11160768 and ANID PIA/APOYO AFB180002.
\end{acknowledgments}


%
%

%


\bibliography{multiscale_stern,nonpolar_solvation,compbio,scbib,bem,fromjay}

\end{document}


\input{Supplementary_material/supp.tex}

%% file: intro.tex
Interactions between a solute molecule and a surrounding solvent are of fundamental importance across chemistry, biology, and associated fields of science and engineering.
Fully atomistic, explicit-solvent molecular dynamics (MD) simulations provide an accurate and chemically detailed understanding of these interactions, but can be prohibitively resource intensive.
Implicit-solvent models~\cite{RouxSimonson1999} require orders of magnitude less computation, but historically their accuracy has been lower than those of explicit-solvent calculations, limiting the scope of possible applications~\cite{Bardhan12_review}.  
Implicit-solvent models have a rigorous statistical-mechanical basis~\cite{RouxSimonson1999}, and can be understood and assessed through the notion of the solvation free energy $\Delta G_{solv}$, the free energy associated with transferring a given solute from gas phase to solvent.  The solvation free energy is frequently decomposed as a sum of two components, the \emph{nonpolar} ($\Delta G_{np}$) and \emph{polar} ($\Delta G_{es}$) terms.
The polar term is the free energy of creating the solute's charge distribution inside a pre-existing solute cavity, and frequently modeled with macroscopic continuum dielectric theory, e.g. the Poisson-Boltzmann equation.

The present paper focuses on the nonpolar component: the work required to place an uncharged solute inside the solvent. Implicit-solvent models often treat $\Delta G_{np}$ as a weighted combination of the solute's solvent-accessible surface area (SASA) and other geometric measures~\cite{GallicchioZhangLevy2002, SwansonHenchmanMcCammon2004, WongAmaroMcCammon2009, WangETal2019}, e.g. $\Delta G_{np} = \gamma (SASA) + b$, where $\gamma$ is usually interpreted as the liquid-vapor surface tension, and $b$ is a scaling constant (sometimes presented as $pV$, pressure times volume).
Nevertheless, the well-known inaccuracies of SASA-based models~\cite{GallicchioKuboLevy2000, WagonerBaker2004, MobleyETal2009,MehdizadehRahimiETal2019} motivate significant research to improve them~\cite{DzubiellaSwansonMcCammon2006, ChengXieDzubiellaMcCammonCheLi2009,NguyenWei2017, WangWangWuWei2018}.
The nonpolar term can itself be decomposed into a sum of two free energies~\cite{Zacharias2003}: $\Delta G_{cav}$, the creation of a solute-shaped cavity by turning on the short-range repulsive interactions (also known as the creation of the \emph{hard-sphere solute}), and $\Delta G_{disp}$, in which one turns on the longer-range attractive dispersion interactions.
In an explicit-solvent description, this is similar to the Weeks-Chandler-Andersen (WCA) theory~\cite{WeeksChandlerAndersen1971}, where one performs MD simulations using a Lennard-Jones potential decomposed into short- and long-range components. Using MD with umbrella sampling, $\Delta G_{cav}$ can also be computed using Lum-Chandler-Weeks (LCW) theory~\cite{LumChandlerWeeks1999,HuangGeisslerChandler2001, VarillyPatelChandler2011}.

The present study on nonpolar solvation is an outgrowth of our longstanding interest in modeling Poisson--Boltzmann (PB) electrostatics using boundary-integral equations~\cite{CooperBardhanBarba2014,Altman06,Altman09,BardhanKnepley2014,TabriziETal2017}. Recognizing that our proposed model is rather unusual, we review this work briefly, to introduce key ideas and motivate our approach.
While using explicit-solvent free-energy pertubation (FEP) MD to parameterize a multiscale PB model~\cite{Bardhan11_pka,Bardhan13_nonlocal_review}, we learned that charge-sign asymmetry errors dominated the differences between standard and multiscale PB~\cite{BardhanJungwirthMakowski2012}, and we identified two distinct sources of charge-sign asymmetry.  
First, water hydrogens can approach solute charges closer than the larger water oxygens.  
The second, central to the present work, is the presence of a significant electrostatic potential in a completely uncharged solute, which is created by the solvent structure immediately surrounding the solute ~\cite{Ashbaugh2000,CeruttiBakerMcCammon2007,Beck13} (see, in particular, Fig. 1 of~\cite{CeruttiBakerMcCammon2007}).  We called this the \emph{static} potential~\cite{Ashbaugh2000}, because it remains when the solute charge distribution is zero everywhere (the familiar \emph{reaction} potential~\cite{RouxSimonson1999} vanishes in the absence of solvent charges).  
Noting that the static potential is nearly constant ($\sim\!10$~kcal/mol/$e$) within even complex molecules~\cite{CeruttiBakerMcCammon2007}, we approximated it as constant and developed a modified dielectric model to account for the steric asymmetry. We called this the SLIC (solvation-layer interface condition) electrostatic model~\cite{BardhanKnepley2014,TabriziETal2017}, and found it sufficiently improved over standard PB that we began to test its performance on total solvation free energies, using the traditional SASA model~\cite{GallicchioZhangLevy2002} for $\Delta G_{np}$.
Our recent study~\cite{MehdizadehRahimiETal2019} showed SLIC/SASA approaching MD in overall accuracy on the Mobley test set~\cite{MobleyETal2009}, despite the fact that the SASA model exhibits a weak overall correlation with explicit-solvent MD results for $\Delta G_{np}$. 
The disparity motivated exploring possible improvements to nonpolar solvation models (others have highlighted this recently as well~\cite{KnightBrooks2011}).

The static potential's existence---and its importance in solvation electrostatics---lead to an apparent paradox: namely, it implies that the solute's nonpolar solvation free energy includes an electrostatic term, which represents the work done creating the static potential inside the solute.  Consider that the static potential field exists in an uncharged solute, i.e., after one creates the solute atoms (as uncharged Lennard-Jones particles, for many FEP calculations).  However, far from the solute, the potential is zero---again, constant. The potential varies significantly only in the solvation shell, indicating a non-zero electric field (and therefore energy density) there.  Therefore, \textit{the thermodynamic steps associated with nonpolar solvation include an electrostatic work}.  The present paper asks, ``How important is this electrostatic work in nonpolar solvation free energies?'' and models the electrostatic problem very simply, as a capacitor in which the molecular solute is treated as a body held at constant potential (the static potential) and the bulk solvent volume treated as a ground (zero potential). We find that this capacitor model reproduces explicit-solvent cavity free energies with surprisingly high accuracy, enabling an overall implicit-solvent model that is comparable in accuracy to explicit-solvent calculations.

%% file: methods.tex
We model the nonpolar solvation free energy $\Delta G_{np} = \Delta G_{cav} + \Delta G_{disp}$, so the solvation free energy is $\Delta G_{solv} = \Delta G_{cav} + \Delta G_{disp} + \Delta G_{es}$. We propose a capacitance-based model for $\Delta G_{cav}$, and combine it with a modification of a continuum integral method for calculating $\Delta G_{disp}$~\cite{FlorisTomasi1989,LevyZhangGallicchioFelts2003,SuleaETal2009}, using SLIC continuum electrostatics~\cite{BardhanKnepley2014,TabriziETal2017,MehdizadehRahimiETal2019} without modification to compute $\Delta G_{es}$.


We model the potential inside the solute as the constant $\phi_{{static}}$, defining the solute boundary as the solvent-excluded surface (SES)~\cite{Connolly83}.
In the infinite solvent region, we model the potential as $\phi = 0$, and bound the region using the boundary of the first solvation shell (the solvent-accessible surface, SAS~\cite{Connolly83}).
These surfaces bound a shell, which we model as a macroscopic dielectric  with the uniform relative permittivity $\epsilon_{shell}$, and where the electrostatic potential obeys the Laplace equation (see SI Figure 1 and derivation in Supporting Information). The fixed potentials at the boundaries mean that the dielectric constants of the solute and bulk solvent are irrelevant.  We use $\epsilon_{shell}$ as a fitting parameter, but it has clear physical bounds and significance.  The energy associated with this boundary-value problem can be solved using a pair of coupled boundary-integral equations for unknown charge densities on the two surfaces,
\begin{equation}\label{eq:matrix}
\left[
\begin{array}{c c}
V_{diel} & V_{diel}\\
V_{shell} & V_{shell}
\end{array}
\right]
\left[
\begin{array}{c}
\sigma_{diel}\\
\sigma_{shell}
\end{array}
\right]
=
\epsilon_{shell}\left[
\begin{array}{c}
\phi_{static}\\
0 
\end{array}
\right],
\end{equation}
where $V_{diel}(\psi_{shell}) = \oint_{shell} \frac{\psi(\mathbf{r}')}{4\pi|\mathbf{r}_{diel} - \mathbf{r}'|}\text{d}\mathbf{r}'$ denotes the potential induced at the inner surface (the dielectric boundary) by the surface charge distribution on the outer boundary (the solvent-accessible surface).  Having found the surface charge distributions, and noting that the potential on the outer surface is zero, we can compute the energy with 
\begin{equation}\label{eq:energy}
\Delta G_{cav} = \oint_{diel} \phi_{static}\sigma_{diel}(\mathbf{r})\text{d}\mathbf{r}.
\end{equation}
Our model therefore treats $\Delta G_{cav}$ as the free energy required to create the interface potential inside the solute.  
This continuum description can also be seen as the free energy of the solvent-solvent electrostatic interactions as molecules accommodate around the cavity, which is related to the so-called reorganization energy~\cite{Lee1985,YuKarplus1988,Lazaridis1998,Lazaridis2000,GallicchioKuboLevy2000} (Supporting Information).
The surface charge layers approximate the behavior of first-shell solvent molecules: when water surrounds a completely uncharged cavity, hydrogens generally point inward, while oxygens are further out~\cite{CeruttiBakerMcCammon2007}.  This creates a kind of diffuse double layer of charge, so $\sigma_{diel}$ roughly captures the water hydrogen charge distribution and $\sigma_{shell}$ represents the water oxygens at the SAS.

We parameterized the molecules with GAFF~\cite{Wang04_GAFF} and discretized the SES and SAS into flat triangular panels (boundary elements) with the \texttt{msms} package \cite{SannerOlsonSpehner1995}, using a $1.4$~\AA-radius probe sphere and taking every atom's radius to be $r_{min}/2$ without adjustment.
We then solved for $\sigma_{diel}$ and $\sigma_{shell}$ with a boundary-element method (BEM) using the \texttt{bempp} library. \cite{SmigajETal2015}
Specifically, we use \texttt{bempp}'s implementation of BEM to generate a finite-dimensional approximation to Equation~\eqref{eq:matrix} by approximating $\sigma_{diel}$ and $\sigma_{shell}$ as piecewise constant on each boundary element, and employing a Galerkin discretization to generate a matrix equation, which we solve using GMRES~\cite{Saad86}. 
We set $\phi_{static}=10.7$~kcal/mol/$e$, following previous work~\cite{BardhanKnepley2014}, and $\epsilon_{shell}=7.75$ (obtained by an empirical single-parameter fitting).

We calculate $\Delta G_{disp}$ by integrating the Lennard-Jones potential outside the solvent-accessible surface~\cite{FlorisTomasi1989,LevyZhangGallicchioFelts2003,SuleaETal2009}, using the divergence theorem to reformulate it as a surface integral 
\begin{equation}\label{eq:vdw_energy}
\Delta G_{disp} = \sum_i \oint_{shell} \rho_w \frac{\partial}{\partial \mathbf{n}}\left( \frac{A_i}{90|\mathbf{r} - \mathbf{r}_i|^{10}}
- \frac{B_i}{12|\mathbf{r} - \mathbf{r}_i|^4}\right) \text{d}\mathbf{r}  
\end{equation}
where $\rho_w$ is the solvent number density (0.0366 \AA$^{-3}$ for water), $A_i$ and $B_i$ are related to the Lennard-Jones parameters for the water oxygen and atom $i$, the sum is over the solute's atoms, and the unit vector $\mathbf{n}$ points into the solvent~\cite{Bardhan07}. 
We found that this model correlates highly with explicit-solvent free energy calculations for $\Delta G_{disp}$, but exhibits a systematic underprediction that correlates highly with surface area.
This continuum Lennard-Jones model treats water density as uniform even though it is known to increase near interfaces~\cite{Beglov96,Beglov97,LakeMcCullagh2017}, so we tested a simple correction in which we allowed water number density in the first solvation shell (outside the SAS) to be higher than bulk; we have used $1.8 \rho_w$ here (see Supporting Information)~\cite{LakeMcCullagh2017}.

In our model for $\Delta G_{np}$, all fitting parameters have clear physical significance, and their optimized values can be at least compared to quantities that are determined relatively easily from experiment or simulation.  For instance, $\epsilon_{shell}$ is between $1$ and $80$ but much closer to 1, with evidence that the permittivity near interfaces depends on the local radius of curvature~\cite{DinpajoohMatyushov2016}.
One can also compute $\phi_{static}$ from explicit-solvent simulations~\cite{CeruttiBakerMcCammon2007,BardhanJungwirthMakowski2012}; for spherical solutes, it varies only slightly with radius~\cite{Ashbaugh2000}.   In this work, the SLIC parameters were $\alpha=0.898$, $\beta=-30.476$, $\gamma=-0.151$, $\mu=-0.449$, and $\phi=0.095$; these were obtained by optimizing SLIC to reproduce explicit-solvent FEP charging free energies~\cite{TabriziETal2017,MehdizadehRahimiETal2019}.


%% file: results.tex

\begin{figure}
    \centering
    \includegraphics[width=0.35\textwidth]{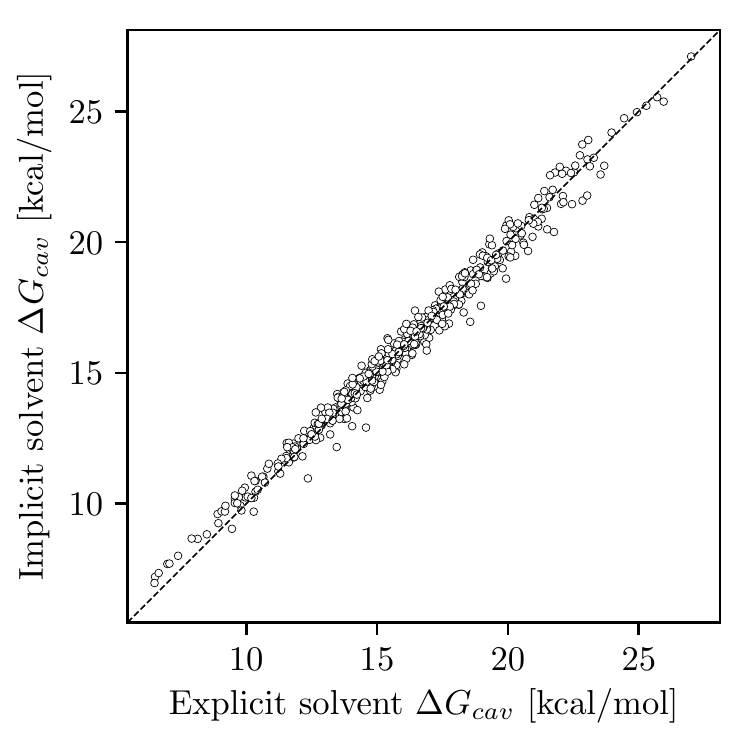}
    \caption{The proposed implicit-solvent model for $\Delta G_{cav}$, compared to explicit-solvent FEP~\cite{MobleyETal2009} ($\rho_p=0.99$, RMSD=$0.45$~kcal/mol, MUE=$0.36$~kcal/mol).}
    \label{fig:dGcav}
\end{figure}

\begin{figure}
    \centering
    \includegraphics[width=0.35\textwidth]{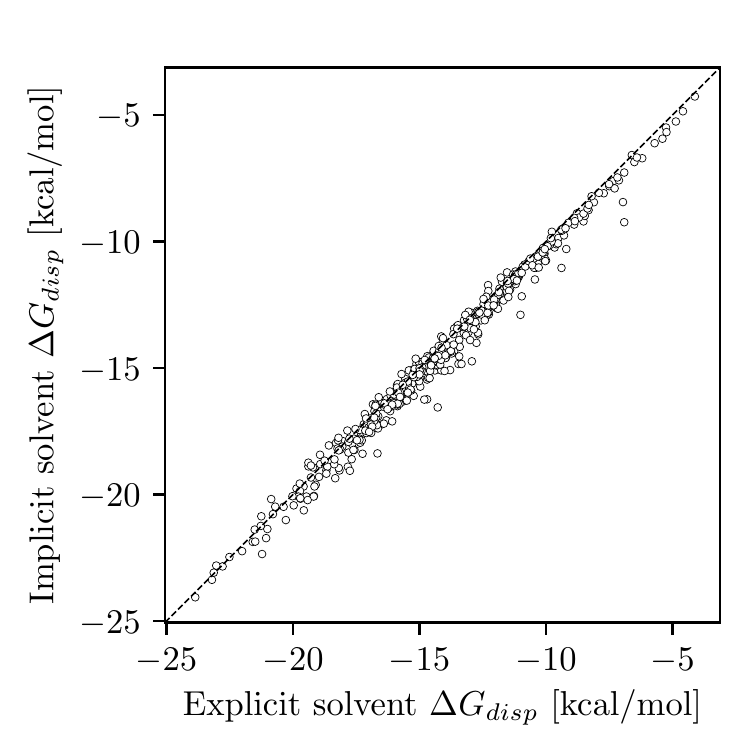}
    \caption{The proposed implicit-solvent model for $\Delta G_{disp}$, compared to explicit-solvent FEP ($\rho_p=0.99$, RMSD=$0.50$~kcal/mol, MUE=$0.38$~kcal/mol).}
    \label{fig:dGdisp}
\end{figure}

\begin{figure}[!h]
\centering
\includegraphics[width=0.35\textwidth]{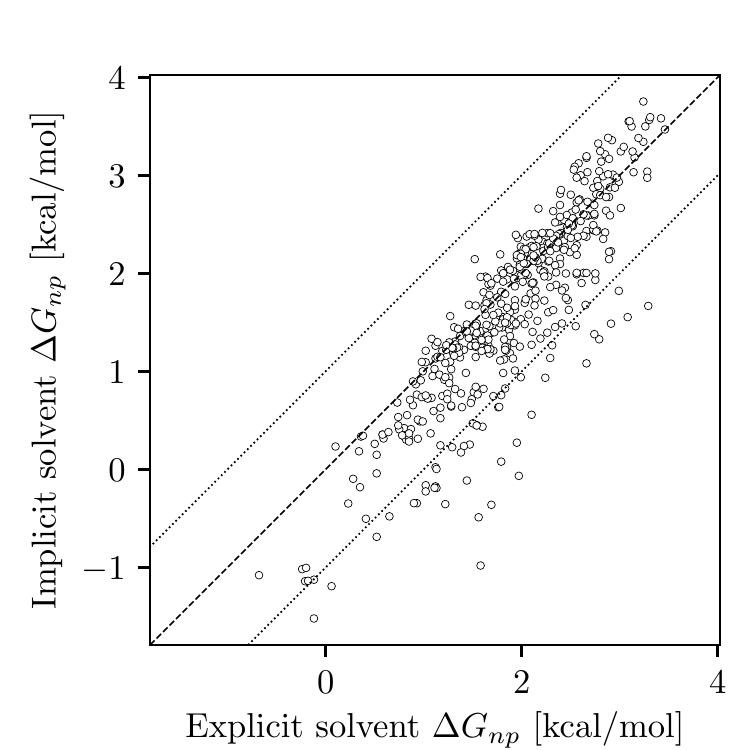}
\caption{The proposed implicit-solvent model for $\Delta G_{np}$, compared to explicit-solvent FEP ($\rho_p=0.88$, RMSD=$0.55$~kcal/mol, MUE=$0.38$~kcal/mol).}
\label{fig:mobley}
\end{figure}

We computed the nonpolar solvation energy of $498$ small solutes from the well-known Mobley test set and compared our results to Mobley's explicit-solvent free-energy perturbation (FEP) calculations~\cite{MobleyETal2009}.
Figures~\ref{fig:dGcav},~\ref{fig:dGdisp}, and~\ref{fig:mobley} are plots of our proposed models for  $\Delta G_{cav}$, $\Delta G_{disp}$, and $\Delta G_{np}$, respectively, compared to Mobley's calculations of the same quantities~\cite{MobleyCavityDispersionDOI}. In each figure, the segmented lines correspond to perfect accuracy; the dotted lines in Figure \ref{fig:mobley} correspond to $\pm\;1$-kcal/mol offsets. We measured accuracy using the Pearson coefficient ($\rho_p$), the root mean square of the difference (RMSD), and the mean unsigned error (MUE); these are given in the figure captions.
The figures show excellent correlation, with $\rho_p=0.88$, RMSD=0.55~kcal/mol, and MUE=0.38~kcal/mol for $\Delta G_{np}$, somewhat lower than for $\Delta G_{cav}$ and $\Delta G_{disp}$, but notable because it is still high despite the fact that $\Delta G_{np}$ is the sum of two large quantities with opposing signs, posing a challenge to model accurately unless the errors are highly correlated.


The proposed model for $\Delta G_{cav}$ explicitly only includes electrostatics, although clearly electrostatics are not the only contribution (Supporting Information). The fact that the model's predictions agree with explicit-solvent calculations indicates that either the electrostatics dominate over other terms in the cavity free energy, or that the dominant components have similar functional forms (e.g. the excellent correlation with SASA~\cite{MobleyETal2009}). 
It is also interesting to note that even though the solvent reorganization energy is usually considered to cancel out in hard-sphere solute solvation~\cite{Lee1985,YuKarplus1988,GallicchioKuboLevy2000}, our model for the associated mean-field electrostatics provides an excellent predictor of cavity free energies. 
This result is supported by previous observations on the importance of the reorganization energy in solvation free energy~\cite{Lazaridis1998,Lazaridis2000}, and reconsiders the role of electrostatics in nonpolar solvation.

\begin{figure}
    \centering
    \includegraphics[width=0.35\textwidth]{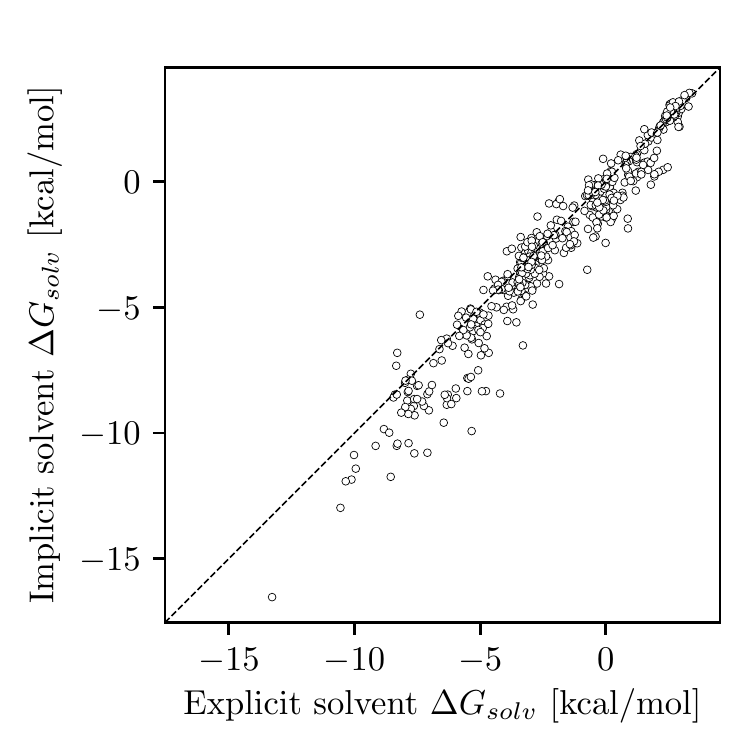}
    \caption{Correlation of $\Delta G_{solv}$ with MD ($\rho_p=0.97$, RMSD=$1.03$~kcal/mol, MUE=$0.72$~kcal/mol).}
    \label{fig:dGtotal}
\end{figure}

Combining SLIC electrostatic calculations on this test set~\cite{BardhanKnepley2014} (SI Figure 4) with our nonpolar model then yields the total $\Delta G_{solv}$.
Figure~\ref{fig:dGtotal} is a plot of the full implicit-solvent model's correlation with explicit-solvent FEP solvation free energies (SI Figure 5 is a plot of the correlation with experiment).  Our implicit-solvent model performs well compared to both explicit-solvent MD results ($\rho_p=0.97$, RMSD=$1.02$~kcal/mol, and MUE=$0.72$~kcal/mol) and experiment ($\rho_p=0.92$, RMSD=$1.35$~kcal/mol, and MUE=$1.01$~kcal/mol).

\begin{figure}
    \centering
    \includegraphics[width=0.35\textwidth]{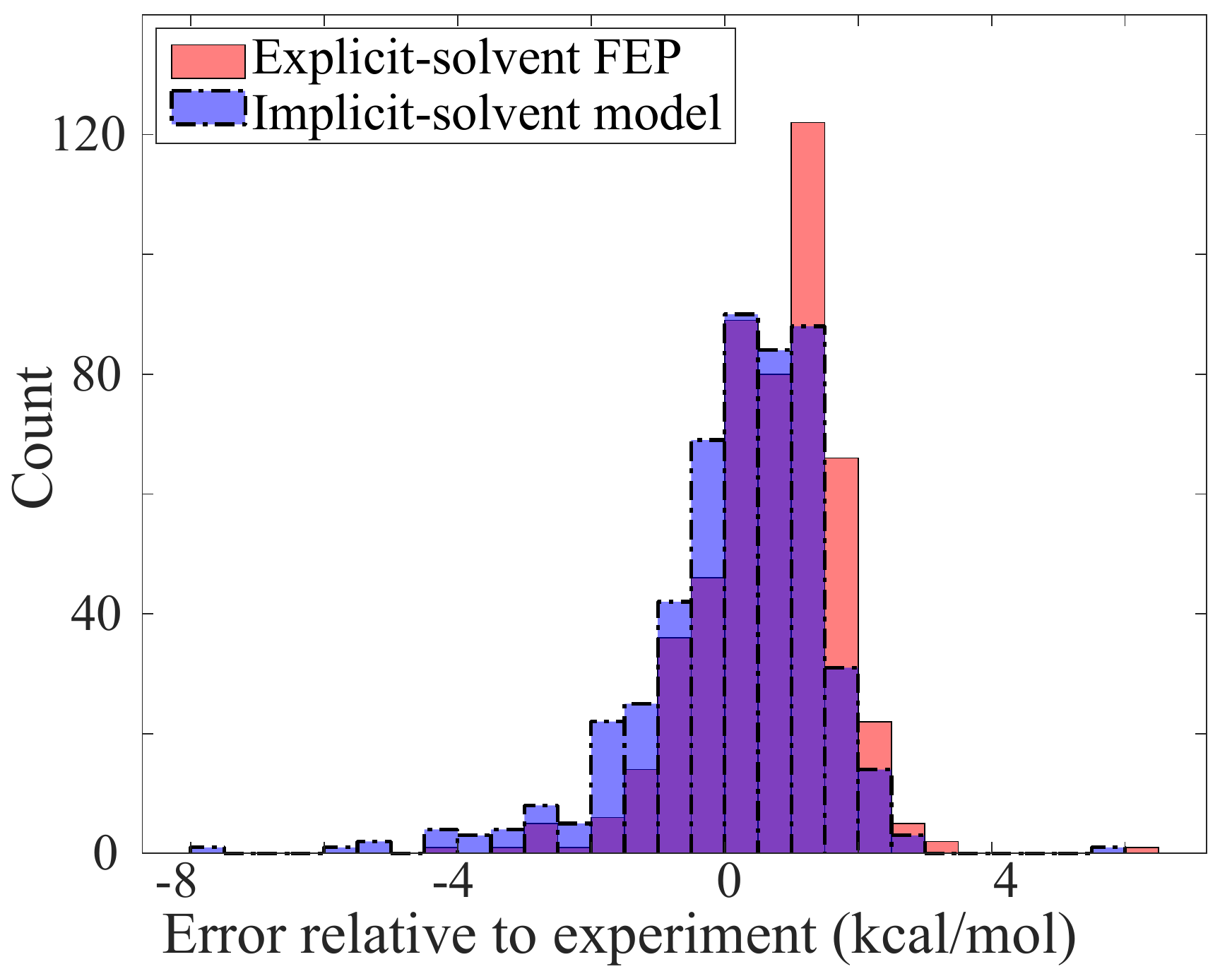}
    \caption{Histograms of errors relative to experiment, for explicit-solvent FEP calculations by Mobley et al., and for the proposed implicit-solvent model ($\rho_p=0.92$, RMSD=$1.35$~kcal/mol, MUE=$1.01$~kcal/mol)}.
    \label{fig:histograms}
\end{figure}

These measures are remarkably comparable to those from explicit-solvent FEP ($\rho_p=0.94$, RMSD=$1.26$~kcal/mol, and MUE=$1.04$~kcal/mol). Beyond these correspondences, the distribution of errors for the implicit-solvent and explicit-solvent models are very similar (Figure~\ref{fig:histograms}), suggesting that physics-based implicit-solvent models may be approaching parity with explicit-solvent MD.

\begin{table}[!h]
\begin{center}
\caption{Comparison with DISA model~\cite{MichaelETal2017}. RMSD and MUE are in kcal/mol.}
\label{table:michael}
\begin{tabular}{c c c c}
Model & $\rho_p$ & RMSD & MUE \\
\hline
DISA  & 0.82  & 0.34 & 0.25 \\
Present work   & 0.92  & 0.27 & 0.20
\end{tabular}
\end{center}
\end{table}

We also compared our proposed nonpolar model to other existing implicit-solvent approaches. 
A recent study~\cite{MichaelETal2017} assessed the performance of different SASA models, using 35 alkanes and 27 polar molecules from the Mobley test set.
The authors obtained the best results using the DISA (dispersion integral surface area) model, which combines a SASA scheme for $\Delta G_{cav}$ and an integral method~\cite{AguilarShadrachOnufriev2010} for $\Delta G_{disp}$. 
Table \ref{table:michael} shows how the DISA model and our approach compare with MD for these solutes.  
Further, the authors found that the most accurate results were obtained when one used a molecule-dependent surface tension---which ranged from $\gamma=-50$ to $\gamma=20$ kcal/mol/\AA, depending on whether the solute was polar, aromatic, an alkane, or an ion~\cite{MichaelETal2017}. Similar dependencies on solute type have been observed previously~\cite{AshbaughKalerPaulaitis1999}.
The dependence of the optimal surface tension on solute type highlights the incomplete physical picture from SASA models:
from the statistical-mechanical perspective~\cite{RouxSimonson1999}, the nonpolar solvation free energy assumes a completely uncharged molecule (a set of Lennard-Jones particles), and therefore a physically consistent model should not depend significantly on atomic charges.
Our model also presents a stronger physical basis for calculating $\Delta G_{disp}$; the \emph{dispersion integral} part of DISA relies on two fitting parameters that to our understanding lack physical interpretation~\cite{AguilarShadrachOnufriev2010}.
In our model for $\Delta G_{disp}$, the only fitting parameters are the shell width and the number density of water molecules near the interface, which is a physical quantity known to be higher than bulk (set here to $1.8\rho_w$).
Similarly, our capacitor approximation for $\Delta G_{cav}$ uses the same parameters ($\epsilon_{shell}$ and $\phi_{static}$) for all the solutes in Mobley's test set, which includes alkane, polar, and aromatic molecules.
Despite the smaller number of free parameters, which are all independent of the solute, our parameterized model outperforms DISA (Table \ref{table:michael}).

The Supporting Information provides additional results, including for spherical solutes, correlations with SASA, and a proof-of-concept calculation of entropy.
In particular, SI Figure 3 shows that SASA correlates linearly with both $\Delta G_{cav}$ and $\Delta G_{disp}$, but exhibits poorer correlation with $\Delta G_{np}$ (SI Figure 5), as observed previously for explicit MD calculations~\cite{MobleyETal2009}. Furthermore, SI Figure 10 shows the notable correlation of $T\Delta S$ computed with our model, compared with experiments, using physically reasonable variations of $\phi_{static}$ and $\rho_w$ with temperature.

%% file: conclusion.tex
The main contribution of this paper is a new approach for estimating the free energy of forming the solute cavity.
Recognizing that the mean electrostatic potential in the (uncharged) cavity is large and approximately constant, and that the mean electrostatic potential outside the solvation layer is zero, we proposed modeling $\Delta G_{cav}$ as a macroscopic capacitor---essentially, the solute-shaped version of a spherical capacitor with two concentric, conducting shells.
From a purely atomistic point of view, the charge structure and electric field in the hydration shell are highly complex.
However, because we were interested in the magnitude of the work required to create the static potential in the uncharged solute, we proposed a very simple model to estimate the total energy stored in the field---neglecting the atomistic details of the charge structure and instead assuming a relatively smooth field.
We found to our surprise that this work corresponded closely to the total cavity free energy.
Our proposed model suggests a previously underappreciated role for electrostatics in solvation: namely, an electrostatic contribution to the Gibbs free energy of cavity formation.

We have modeled $\Delta G_{disp}$ with a modified integral-type continuum model, in which one sums over all solute atoms the volume integral of the atom's Lennard-Jones interactions with the solvent. We found that our implementation of the standard integral approach systematically underpredicted $\Delta G_{disp}$.  We improved the predictions significantly by allowing higher-than-bulk solvent density in the hydration shell, in accordance with experimental and simulation results; previous such models assumed the solvent density to be bulk everywhere. Given the accuracy of WCA nonpolar energies~\cite{WagonerBaker2006}, efforts to improve WCA models may benefit from similarly granular comparisons. 

We have validated the overall model (capacitance-based cavity free energy, modified continuum integral dispersion, and SLIC electrostatics) against explicit-solvent MD free-energy calculations and found that it predicts nonpolar solvation free energies more accurately than models based on the common solvent-accessible surface-area (SASA) model.
Importantly, our model achieves this accuracy without employing individual atomic radii as fitting parameters in any of the three energy terms.
The nonpolar and the electrostatic terms use the same underlying atomic radii from the MD force field ($r_{min}/2$), with a uniform scaling for SLIC electrostatics such that for any atom $r_{implicit} = 0.92 r_{min}/2 $.
This uniform scaling factor of 0.92 was set in our first study of the SLIC continuum electrostatic model~\cite{BardhanKnepley2014}, and we have not attempted to reparameterize or find a different scaling factor since the original paper.
The parameters of our proposed model (static potential, permittivity, and solvation-shell number density) have robust physical interpretations that are largely intrinsic to the solvent, and do not depend on solute type or atom types.
We highlight this aspect of our model because the thermodynamic cycles often drawn to distinguish contributions to the nonpolar solvation free energy involve processes in which the associated models should not depend strongly on atom type or solute size. We acknowledge that our model's parameters do have a minor dependence, see e.g.~\cite{Ashbaugh2000} for the variation in static potential with solute size, but even as a fitting parameter, it is one that works for all solute types tested so far.
Also, although the precise values used in the proposed model may be viewed to some extent as fitted parameters, the model gives accurate results for physically reasonable values.
The correlation of such a simple model with explicit-solvent MD seems notable, especially considering the small number of fitting parameters.

We have also found that our capacitance-based model for $\Delta G_{cav}$ exhibits a temperature dependence that correlates surprisingly well with solvation entropies when one uses a reasonable value for the temperature dependence of the static potential.  Although the combination of SLIC electrostatics and the SASA nonpolar model reproduces solvation entropies well on available data, doing so requires a grossly unphysical change in surface tension with temperature~\cite{MehdizadehRahimiETal2019}. This is intriguing because $\Delta G_{cav}$ and $\Delta G_{disp}$ correlate well with the solvent-accessible surface area (SASA), but
much more poorly with the total $\Delta G_{np}$~\cite{MobleyETal2009,MehdizadehRahimiETal2019} (SI Figure 9).  Overall, the proposed model appears to represent progress for implicit-solvent theory.  Not only do the individual terms $\Delta G_{cav}$ and $\Delta G_{disp}$ correlate well with explicit-solvent results using physically reasonable parameters, they also (1) sum to an accurate model for $\Delta G_{np}$; and (2) correlate well with solvation entropies, given known temperature dependencies~\cite{Jones92,Beck13}. The significance of this correlation is a subject of current work.

%% file: Supplementary_material/supp.tex
\title{Supplementary material for ``A Simple Electrostatic Model for the Hard-Sphere Solute Component of Nonpolar Solvation''}
\author{Christopher D. Cooper and Jaydeep P. Bardhan}
\maketitle

\section{The solvation energy}
\begin{figure}[!h]
\centering
\includegraphics[width=0.8\textwidth]{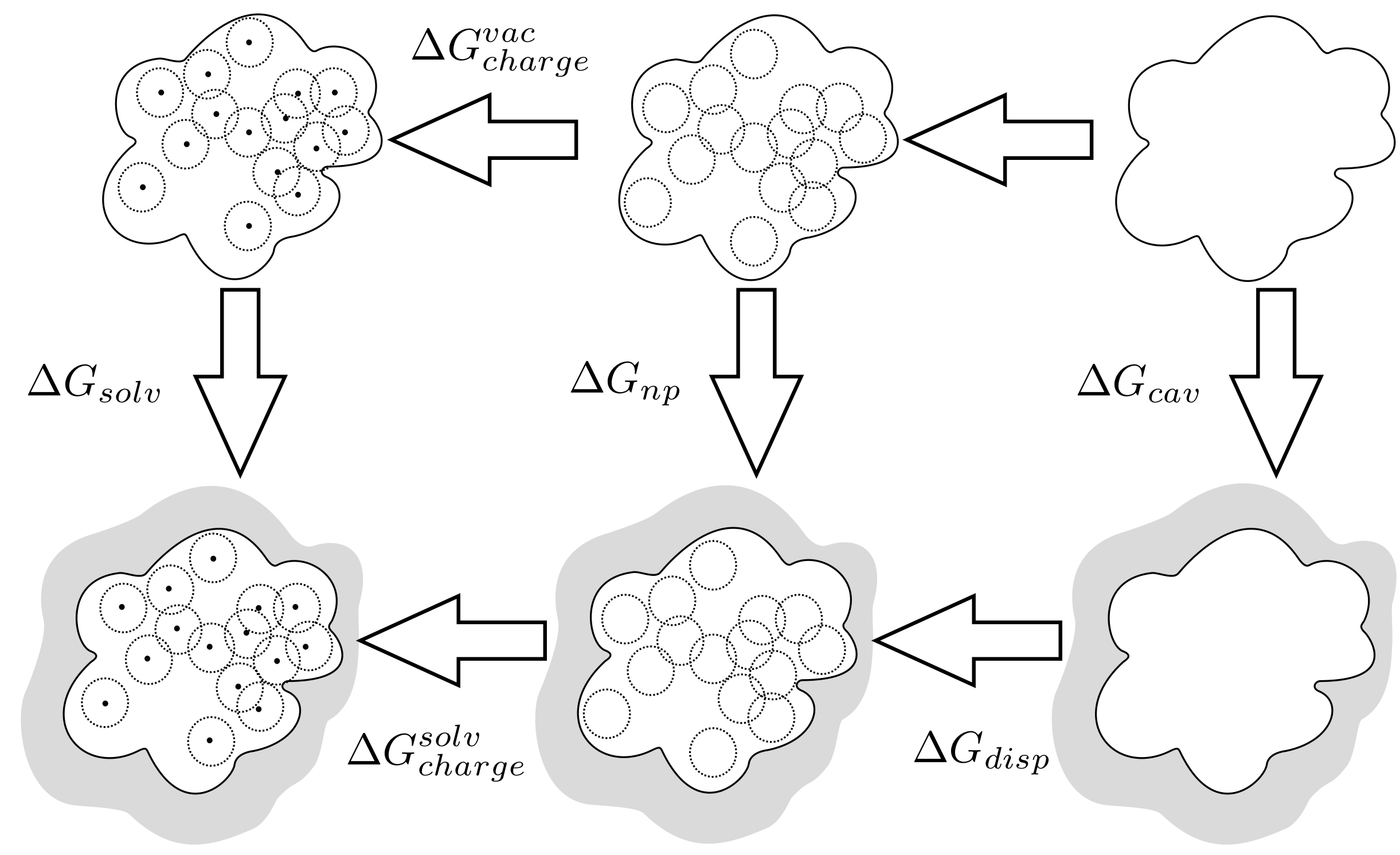}
\caption{Thermodynamic cycle for solvation energy.}
\label{fig:thermo_cycle}
\end{figure}

Following the work by Levy and co-workers~\cite{LevyZhangGallicchioFelts2003}, the solvation energy can be computed using the thermodynamic cycle from Figure \ref{fig:thermo_cycle}.
In it, the solvation energy ($\Delta G_{solv}$) is decomposed into polar ($\Delta G_{es}$) and non-polar ($\Delta G_{np}$) parts.
The polar component is the difference in charging energy between vaccum and solvated states 
\begin{equation}\label{eq:G_polar}
\Delta G_{es} = \Delta G_{charge}^{solv} - \Delta G_{charge}^{vac}, 
\end{equation}
whereas the nonpolar part is the sum of the energy required to create the cavity in the solvent and the energy from the solvent-solute dispersion interaction
\begin{equation}\label{eq:G_nonpolar}
\Delta G_{np} = \Delta G_{cav} + \Delta G_{disp}.
\end{equation}

The well-known WCA theory~\cite{WeeksChandlerAndersen1971} divides the Lennard-Jones 6---12 potential into repulsive \emph{short-range} and attractive \emph{long-range} terms 
\begin{align}\label{eq:wca}
&\phi_{rep}(r) = \left\{ 
\begin{array} {l l}
\phi_{LJ} + \epsilon & r\leq r_{min}/2\\
0 & r>r_{min}/2
\end{array} \right. \nonumber\\
&\phi_{att}(r) = \left\{ 
\begin{array} {l l}
-\epsilon & r\leq r_{min}/2\\
\phi_{LJ} & r>r_\text{min}/2
\end{array} \right. 
\end{align}
Relating Equation \eqref{eq:G_nonpolar} with the WCA theory, $\Delta G_{cav}$ approximately corresponds to the hard-sphere solute free energy (considering only $\phi_{rep}$), and $\Delta G_{disp}$ is roughly equivalent to the free energy from $\phi_{att}$.


\section{A dielectric interface with a surface charge distribution}
\input{Supplementary_material/appendix.tex}

\section{A statistical mechanical interpretation of the cavity energy}

We interpret $\Delta G_{cav}$ from a statistical mechanics perspective as a reorganization free energy, in which the cavity creation occurs in a two-step thermodynamic process. We first create a solute-shaped hard-sphere solute in an ideal-gas like solvent (a repulsive Lennard-Jones fluid)  by growing the solute's short-range Lennard-Jones potential (the near-field term of WCA). 
In the second step, we turn on the solvent molecules' charges and the long-range dispersion interactions.

The work in the first step is done against pressure ($p\text{d}V$).  The free energy in the second step represents the work done by including solvent-solvent interactions, as the molecules accommodate around the solute cavity; this is known as the solvent reorganization energy~\cite{Lee1985,YuKarplus1988,Lazaridis1998,Lazaridis2000,GallicchioKuboLevy2000}.
The Gibbs free energy change equals the work performed on a system at constant pressure and temperature, minus the pressure contribution ($p\text{d}V$), hence, the change in Gibbs free energy in the first step should be zero.
On the other hand, the solvent reorganization energy in the second step has two components: electrostatic and dispersive. 
Our capacitor model is a continuum electrostatic description of this solvent-solvent interaction that entails averaging over the solvent degrees of freedom, making Equation 2 of the main text a free energy~\cite{RouxSimonson1999}.
In that sense, $\Delta G_{cav}$ models the electrostatic component of the solvent reorganization \emph{free} energy, as the solvent molecules accommodate from bulk (in completely random orientations) to arrangements surrounding the cavity.

In our model, both the enthalpic and entropic terms of the Gibbs free energy are being considered implicitly ($\Delta G = \Delta H + T\Delta S$), and should be able to capture, for example, enthalpy-entropy compensation in nonpolar solvation~\cite{Lee1985,YuKarplus1988,GallicchioKuboLevy2000}.
We can recast the entropic component by studying the relation of $\phi_{static}$, $\epsilon_{shell}$, and $\rho_w$ with temperature ($T$), and computing entropy as $\Delta S = \partial \Delta G_{np}/\partial T$, which would be useful to understand the contribution of entropy in the hydrophobic effect.
As a first attempt, we present results along these lines in this Supplementary Information.

\section{Results}
\subsection{Cavity energy for a sphere}

\begin{figure}[!h]
\centering
\includegraphics[width=0.5\textwidth]{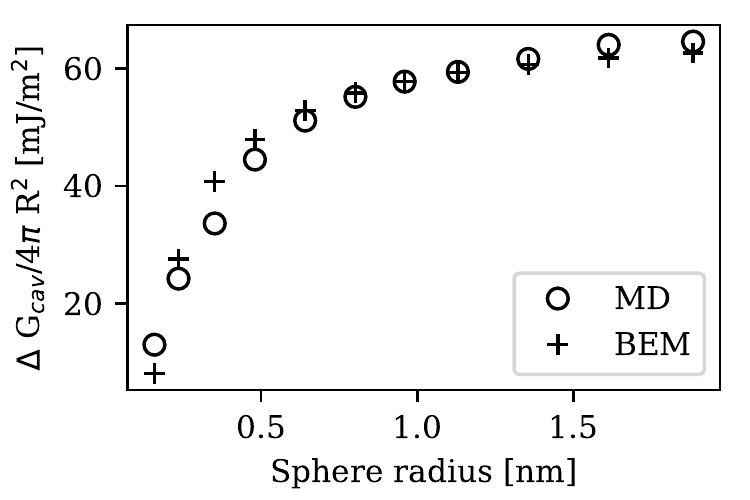}
\caption{$\Delta G_{cav}$ per unit surface area for a sphere using our model (BEM), compared to explicit solvent calculations taken from the work of Chandler \cite{Chandler2005}.}
\label{fig:sphere}
\end{figure}

As an initial test, we computed the solvation energy of a hard sphere with our method ($\Delta G_{cav}$), and compared it with explicit solvent calculations from Huang {\it et al.} \cite{HuangGeisslerChandler2001}, which are also available in a review article by David Chandler \cite{Chandler2005}.
The results from Figure \ref{fig:sphere} were obtained using $\phi_{static}=10$~kcal/mol/$e$, and a permittivity of $\epsilon_{stern}=18$ in the Stern layer.

The $y$ axis in Figure \ref{fig:sphere} is the cavity energy per unit surface area, and it shows a trend where it initially grows with the sphere radius, and then asymptotes to a constant value.
This behavior, in which the cavity free energy of a small sphere scales with volume, and then with surface area, has been widely discussed \cite{Chandler2005}, and our capacitor model captures it naturally. 

\subsection{The solvation-layer interface condition (SLIC) for electrostatics}
\begin{figure}[!h]
\centering
\includegraphics[width=0.5\textwidth]{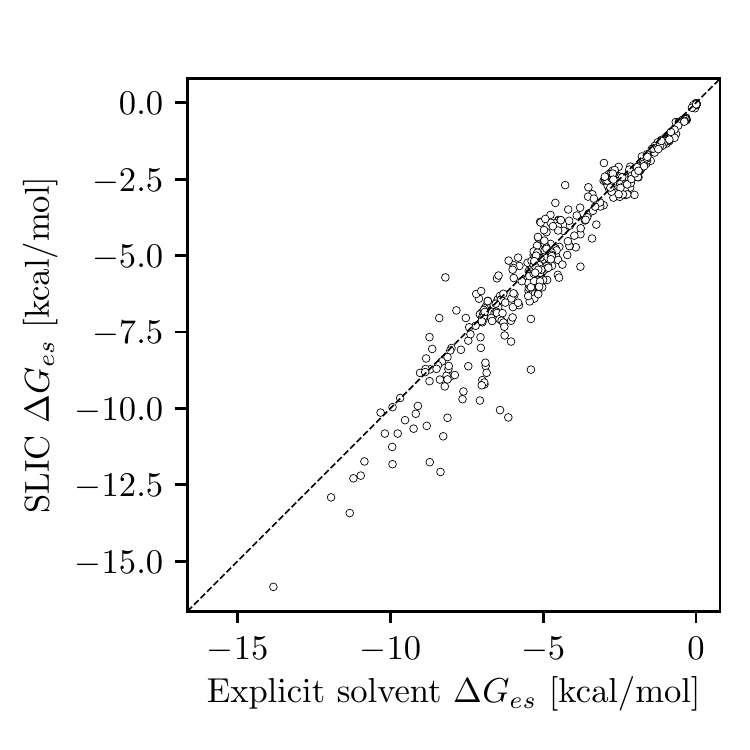}
\caption{Correlation of $\Delta G_{es}$ between SLIC and MD-FEP~\cite{MobleyETal2009} ($\rho_p$=0.97, RMSD=0.72 kcal/mol, MUE=0.47 kcal/mol).}
\label{fig:dGes}
\end{figure}

We used the SLIC model for the electrostatic component~\cite{BardhanKnepley2014}, which applies a modified interface condition on the molecular surface that is capable of capturing charge asymmetry in solvation.
Figure \ref{fig:dGes} shows the correlation of $\Delta G_{es}$ of SLIC with respect to the explicit-solvent FEP charging free energies of Mobley~\cite{MobleyETal2009}.

\begin{figure}[!h]
\centering
\includegraphics[width=0.5\textwidth]{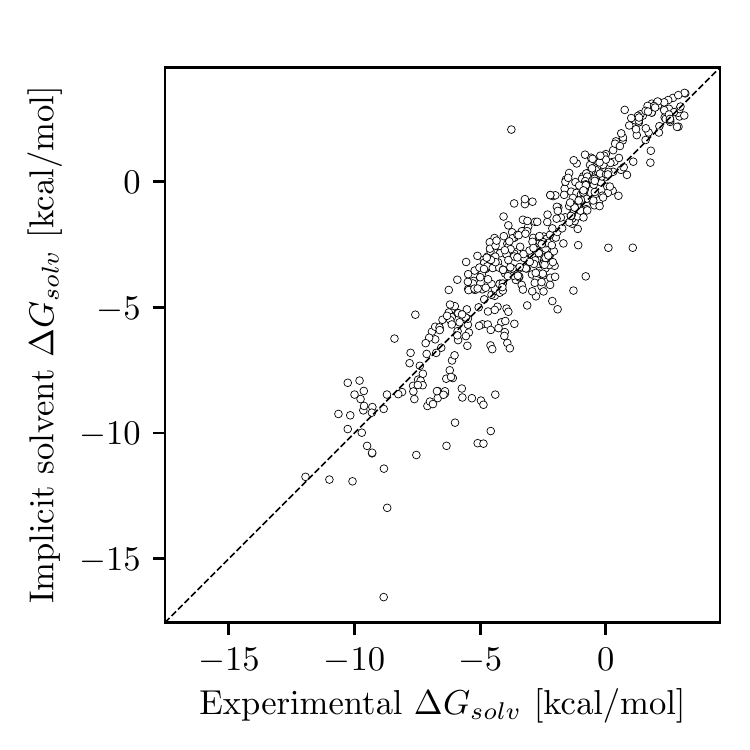}
\caption{$\Delta G_{solv}$ with our approach compared to experiments~\cite{MobleyETal2009} ($\rho_p$ = 0.92, RMSD=1.35 kcal/mol, MUE=1.01 kcal/mol).}
\label{fig:dGexp}
\end{figure}

Having $\Delta G_{es}$, we can compute the total solvation free energies $\Delta G_{solv} = \Delta G_{cav} + \Delta G_{disp} + \Delta G_{es}$, and compare it with experiment (Figure \ref{fig:dGexp}). 

\begin{figure}[!h]
\centering
\includegraphics[width=0.5\textwidth]{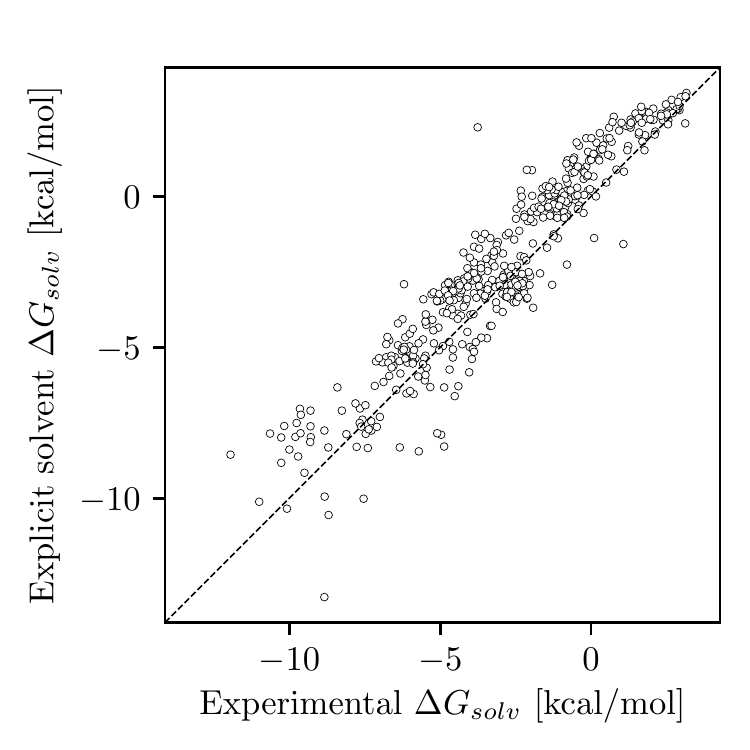}
\caption{$\Delta G_{solv}$ computed with MD compared with experiments~\cite{MobleyETal2009} ($\rho_p$ = 0.94, RMSD=1.26 kcal/mol, MUE=1.04 kcal/mol).}
\label{fig:dGexpMD}
\end{figure}

To illustrate the similarities between our proposed implicit-solvent model and the results from explicit-solvent FEP, Figure~\ref{fig:dGexpMD} is the analogous correlation plot of the FEP-calculated $\Delta G_{solv}$ compared to experiment~\cite{MobleyETal2009}.  Furthermore, the correlation statistics between the explicit-solvent and implicit-solvent models are quite close (see the main text), possibly indicating that further improvements would require improving the underlying force field used as a reference for parameterizing our implicit-solvent model.

\subsection{Correlation with solvent-accessible surface area}

\begin{figure}[!h]
\centering
\includegraphics[width=0.45\textwidth]{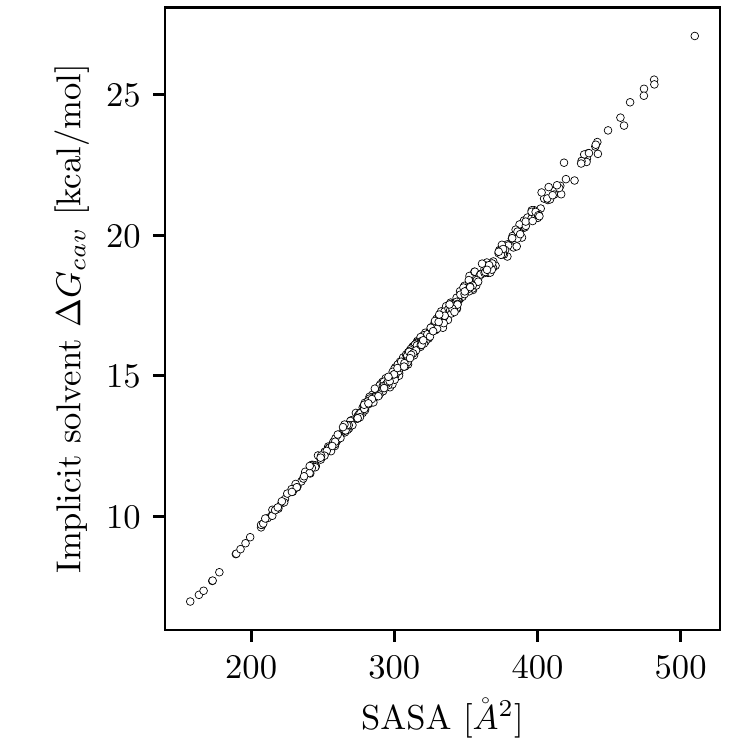}
\includegraphics[width=0.45\textwidth]{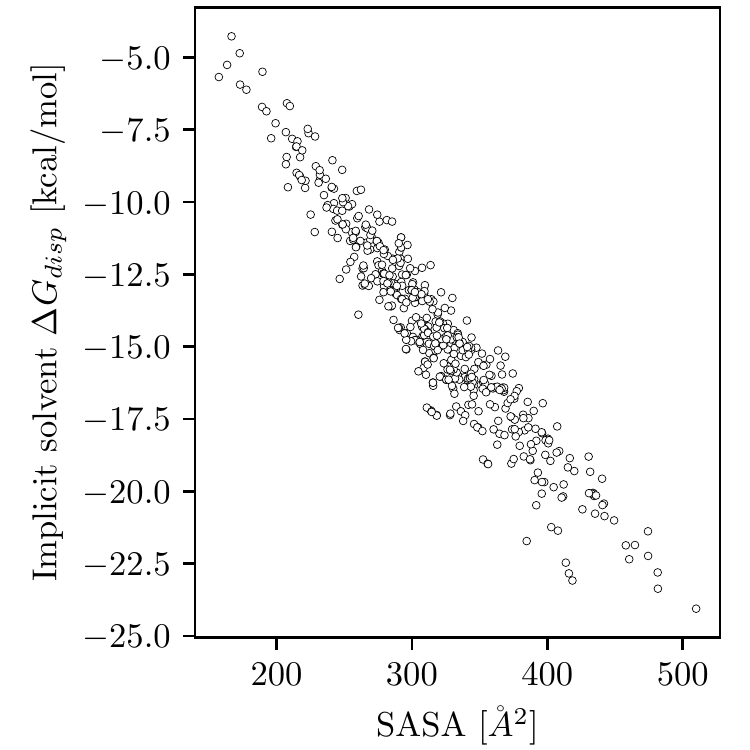}
\caption{Correlation of $\Delta G_{cav}$ and $\Delta G_{disp}$ with solvent-accessible surface area}
\label{fig:area_correl}
\end{figure}

\begin{figure}[!h]
\centering
\includegraphics[width=0.45\textwidth]{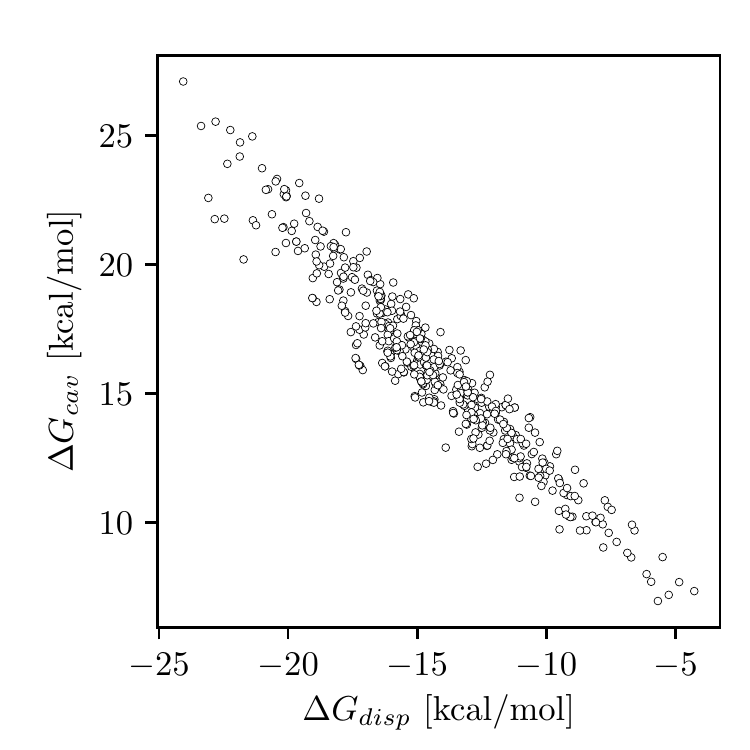}
\caption{Correlation of $\Delta G_{cav}$ with $\Delta G_{disp}$}
\label{fig:cav_vdw_correl}
\end{figure}

\begin{figure}[!h]
\centering
\includegraphics[width=0.45\textwidth]{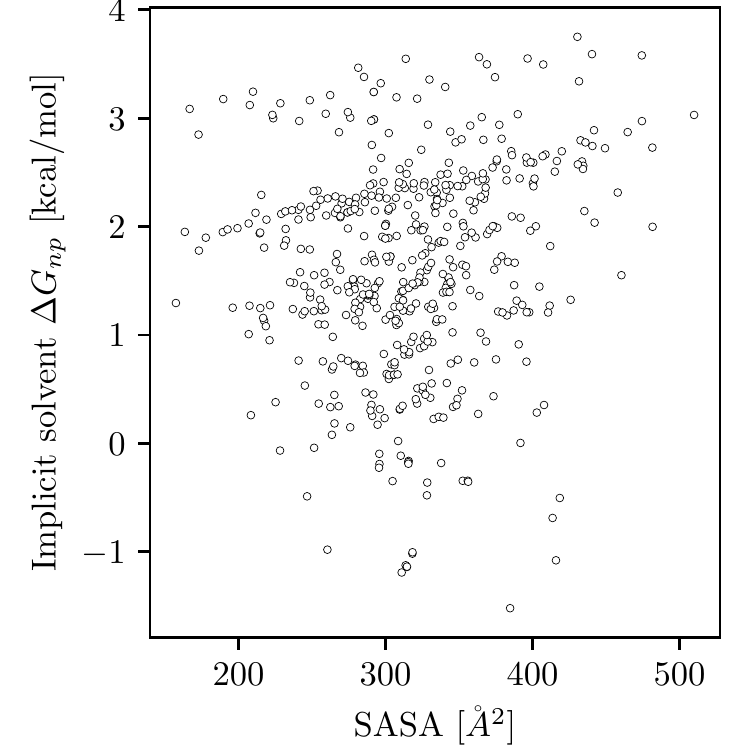}
\caption{Correlation of $\Delta G_{np}$ with solvent-accessible surface area}
\label{fig:area_correl_full}
\end{figure}

The molecular dynamics calculations in Mobley {\it et al.} \cite{MobleyETal2009} show that area is positively correlated with the solvation energy of a hard-sphere solute (only with $\phi_{rep}$ in Equation \eqref{eq:G_nonpolar}), and negatively correlated with the energy from the solute-solvent dispersion ($\phi_{att}$ in Equation \eqref{eq:G_nonpolar}).  As a consequence, the two contributions to $\Delta G_{np}$ are themselves negatively correlated.   The total nonpolar energy, however, correlates poorly with surface area. Our model gives similar results for the correlation of SASA with the two nonpolar terms we have proposed (Figure \ref{fig:area_correl}), that is, with the cavity energy computed using our capacitor model, and the dispersion with the solvation-shell modified surface integral approach from Bardhan and co-workers \cite{BardhanETal2005}.  In Figure \ref{fig:area_correl}, the SASA strongly correlates with $\Delta G_{cav}$ (coefficient of $0.999$), and also with $\Delta G_{disp}$ (coefficient of $-0.95$).  Similar to Mobley's findings, the correlation coefficient between them is $-0.96$ (Figure \ref{fig:cav_vdw_correl}), and the total nonpolar solvation free energy correlates poorly (correlation coefficient $0.22$, see 
Figure \ref{fig:area_correl_full}).

The correlation of SASA with the individual terms provides one justification for the popularity of SASA-based models for calculating $\Delta G_{cav}$ and often $\Delta G_{np}$. However, beyond the notion of a surface tension, these fail to give any physical insight of the cavity formation process, and there are clearly weaknesses associated with trying to use a SASA model for the total $\Delta G_{np}$.  On the other hand, our capacitance model exhibits the same apparent correlation between a solute's SASA and its $\Delta G_{cav}$, which is not unexpected. For a capacitor composed of two parallel plates with area $A$, a distance $d$ apart, which are separated by a dielectric of relative permittivity $\epsilon$: the energy when one is grounded and the other is at a potential $V$ is given by $E = \frac{1}{2} \frac{A}{\epsilon d} V^2$.

The total nonpolar solvation free energy is the result of cancellation (competition) between an unfavorable $\Delta G_{cav}$ and a favorable $\Delta G_{disp}$, and their sum $\Delta G_{np}$ may be one order of magnitude smaller than the magnitudes of the two components.  This fact amplifies the need to have accurate physical models for the components, because relatively small deviations in either component become magnified in the total $\Delta G_{np}$.  The SASA approaches give a clear example of this challenge: each term correlates strongly with SASA, but their sum does not correlate strongly at all.  The large cancellation between $\Delta G_{cav}$ and $\Delta G_{disp}$ also has an important implication for calculations: to obtain an accurate total $\Delta G_{np}$, it is essential to use highly accurate numerical methods for the independent terms (in order to avoid having numerical errors that exceed the energetic difference, which is the quantity of main interest).

Last, we wish to highlight a particular strength of our model.  For most solutes in this test set, the cavity energy is slightly larger in magnitude than the dispersion energy, and therefore the total nonpolar solvation energy is positive. However, there are cases where the total nonpolar solvation free energy is negative; in the Mobley test set, we have observed that this seems to happen more often for some nitrogen-containing compounds, such as 4-nitrophenol.
These unusual, negative nonpolar solvation free energies are well captured by our model.







\subsection{Entropy calculations}
As an initial exploration of our model's implications for the entropic component of $\Delta G_{solv}$, we computed Gibbs solvation free energies at two temperatures, equally above and below 300~K (the baseline temperature for calculations), approximating $T\Delta S$ with a central difference as
\begin{equation}\label{eq:tds}
T\Delta S = T \frac{\partial \Delta G_{solv}}{\partial T} \approx T \frac{\Delta G_{solv}^{over} - \Delta G_{solv}^{under}}{\Delta T}.
\end{equation}
The influence of temperature on $\rho_w$ and $\phi_{static}$ is reasonably well understood~\cite{Jones92,Beck13} but its effect on $\epsilon_{shell}$ remains unclear~\cite{DinpajoohMatyushov2016,MostajabiSarhangi19}, so we left it unchanged.
In previous studies of small-molecule solvation thermodynamics with the SLIC electrostatic model and SASA nonpolar model~\cite{MehdizadehRahimiETal2019}, we have found that our parameterized $\Delta G_{es}$ varies only slightly over this temperature range (unpublished results). As a result, our predictions of entropies consider only the effects of temperature variations in $\phi_{static}$ and $\rho_w$. Specifically, we assumed that an 8~K increase in temperature led to an increase of 0.06~kcal/mol/$e$ in $\phi_{static}$, and a decrease of 0.3\% in $\rho_w$.
Thus, at 292 K, we used $\phi_{static}=10.64$ kcal/mol/e and $\rho_w$=0.0337 \AA$^{-3}$ to compute $\Delta G_{solv}^{under}$, and at 308 K $\phi_{static}=10.76$ kcal/mol/e and $\rho_w$=0.0335 \AA$^{-3}$ for $\Delta G_{solv}^{over}$.  Importantly, these parameter variations are close to observed changes.  Beck observed, for SPC/E water, an 0.3~kcal/mol/$e$ increase in the static potential over 50~K~\cite{Beck13}. Experimental data regarding the density variation of bulk water show a decrease of approximately 0.6\% from 290~K to 310~K~\cite{Jones92}.

\begin{figure}[!h]
\centering
\includegraphics[width=0.45\textwidth]{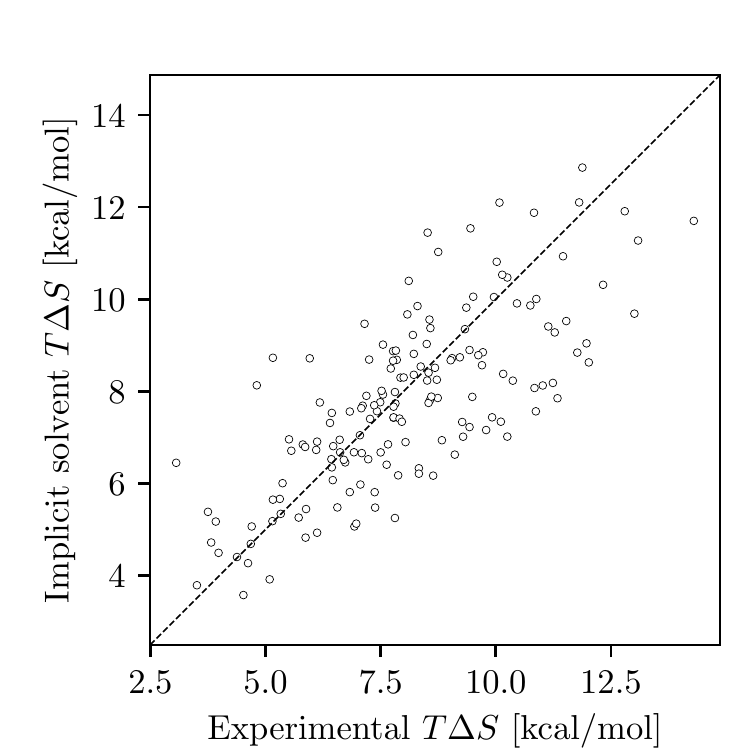}
\caption{Comparison of $T\Delta S$ between our approach and experiments ($\rho_p$ = 0.75, RMSD=1.49 kcal/mol, MUE=1.16 kcal/mol).}
\label{fig:TdS_comparison}
\end{figure}

Figure \ref{fig:TdS_comparison} compares $T\Delta S$ computed using Equation \eqref{eq:tds} with experiments~(tabulated by A. Mehdizadeh Rahimi, manuscript in preparation), on a subset of 159 of Mobley's solutes (the experimental data were originally from~\cite{Cabani81,Barone90,Makhatadze93_2,Plyasunov99,Plyasunova05}).  Considering our model's simplicity and our use of parameter variations comparable to experimental observations, the results correlate remarkably well ($\rho_p$ = 0.75, RMSD=1.49 kcal/mol, MUE=1.16 kcal/mol). This highlights the physical foundation behind our model, in contrast to SASA approaches, where unphysical changes in the surface tension are required to correctly reproduce $T\Delta S$~\cite{MehdizadehRahimiETal2019}.
Thus it is surprising to see that, apparently, there is an important electrostatic component in hydrophobicity, or at least, it yields an outstanding predictor.

\bibliographystyle{abbrv}
\bibliography{multiscale_stern,nonpolar_solvation,compbio,scbib,bem,fromjay}

%% file: Supplementary_material/appendix.tex
\begin{figure}
\centering
\includegraphics[width=0.4\textwidth]{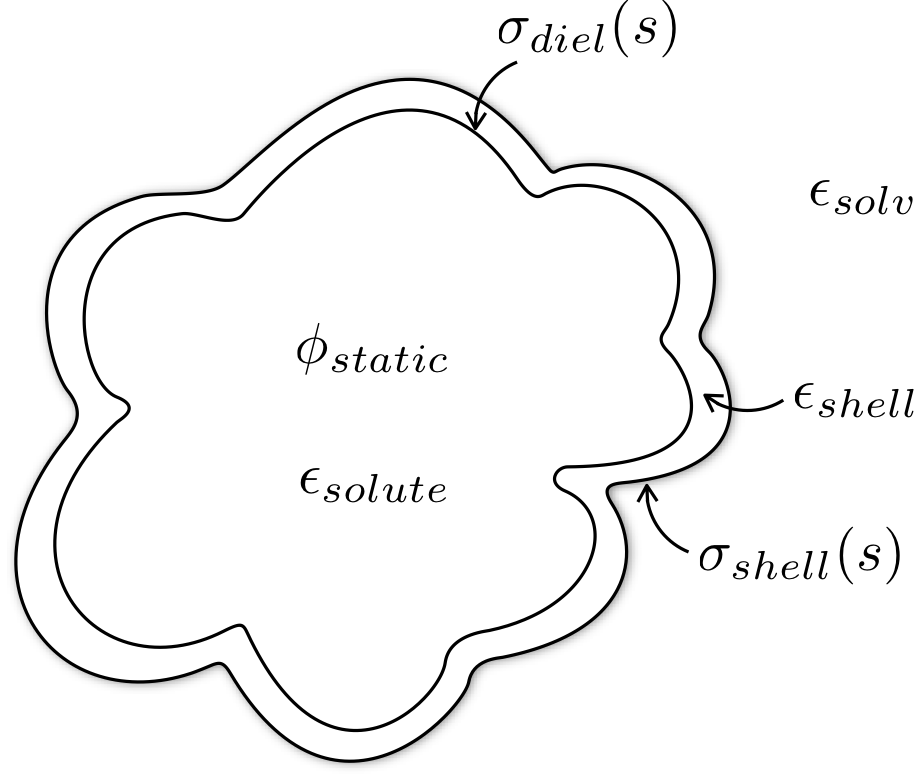}
\caption{Sketch of the capacitor model}
\label{fig:capacitor}
\end{figure}

Applying continuum electrostatic theory on the system in Figure \ref{fig:capacitor} yields the following partial differential equations
\begin{align}\label{eq:pde}
\nabla^2 \phi_{solute} &= 0 \text{ on $\Omega_{solute}$},\nonumber\\
\nabla^2 \phi_{shell} &= 0 \text{ on $\Omega_{shell}$},\nonumber\\
\nabla^2 \phi_{solv} &= 0 \text{ on $\Omega_{solv}$},\nonumber\\
\phi_{solute} = \phi_{shell} &\text{ and } \epsilon_{solute}\frac{\partial \phi_{solute}}{\partial \mathbf{n}} - \epsilon_{shell}\frac{\partial \phi_{shell}}{\partial \mathbf{n}} = \sigma_{diel}\text{ at $\Gamma_{diel}$}, \nonumber\\
\phi_{prot} = \phi_{solv} &\text{ and } \epsilon_{shell}\frac{\partial \phi_{shell}}{\partial \mathbf{n}} - \epsilon_{solv}\frac{\partial \phi_{solv}}{\partial \mathbf{n}} = \sigma{shell}\text{ at $\Gamma_{shell}$}
\end{align}
which can be represented in integral form as
\begin{align}\label{eq:integral_rep}
&\int_{\Omega_{solute}} \phi_{solute}(\mathbf{r}')\delta(|\mathbf{r}-\mathbf{r}'|){d}\mathbf{r}' = V_\mathbf{r}^{\Gamma_{shell}}\left(\frac{\partial\phi_{solute}}{\partial\mathbf{n}}\right) - K_\mathbf{r}^{\Gamma_{diel}}(\phi_{solute})\nonumber \\
&\int_{\Omega_{shell}} \phi_{shell}(\mathbf{r}')\delta(|\mathbf{r}-\mathbf{r}'|){d}\mathbf{r}' = -V_\mathbf{r}^{\Gamma_{diel}}\left(\frac{\partial\phi_{shell}}{\partial\mathbf{n}}\right) + K_\mathbf{r}^{\Gamma_{diel}}(\phi_{shell}) + V_\mathbf{r}^{\Gamma_{shell}}\left(\frac{\partial\phi_{shell}}{\partial\mathbf{n}}\right) - K_\mathbf{r}^{\Gamma_{shell}}(\phi_{shell})\nonumber \\
&\int_{\Omega_{solv}} \phi_{solv}(\mathbf{r}')\delta(|\mathbf{r}-\mathbf{r}'|){d}\mathbf{r}' = - V_\mathbf{r}^{\Gamma_{shell}}\left(\frac{\partial\phi_{solv}}{\partial\mathbf{n}}\right) + K_\mathbf{r}^{\Gamma_{shell}}(\phi_{solv})
\end{align}
where $K_\mathbf{r}^{\Gamma}(\psi)$ is the double layer potential of distribution $\psi$ on $\Gamma$, evaluated at $\mathbf{r}$:
\begin{equation}
K_\mathbf{r}^\Gamma (\psi) = \oint_\Gamma \psi(\mathbf{r}') \frac{\partial}{\partial\mathbf{n}}\left(\frac{1}{4\pi|\mathbf{r}-\mathbf{r}'|}\right)\text{d}\mathbf{r}'
\end{equation}
and $V_\mathbf{r}^\Gamma (\psi)$ the single layer potential from Equation (1) in the main text.

If we evaluate $\mathbf{r}$ in the shell region ($\Omega_{shell}$), the left hand side of Equation \eqref{eq:integral_rep} is zero for the integrals over $\Omega_{solute}$ and $\Omega_{solv}$, and $\phi_{shell}$ for the middle one.
Adding the three equations, and cancelling out all double-layer terms considering that the potential is continuous accross the interfaces, we arrive to
\begin{equation}
\phi_{shell} = V_{\Omega_{shell}}^{\Gamma_{diel}}\left(\frac{\partial\phi_{prot}}{\partial\mathbf{n}}\right) - V_{\Omega_{shell}}^{\Gamma_{diel}}\left(\frac{\partial\phi_{shell}}{\partial\mathbf{n}}\right) + V_{\Omega_{shell}}^{\Gamma_{shell}}\left(\frac{\partial\phi_{shell}}{\partial\mathbf{n}}\right) - V_{\Omega_{shell}}^{\Gamma_{shell}}\left(\frac{\partial\phi_{solv}}{\partial\mathbf{n}}\right).
\end{equation}
Applying the interface conditions on the electric displacement from Equation \eqref{eq:pde} yields
\begin{align}\label{eq:phi_stern}
\phi_{shell} = &\left(1-\frac{\epsilon_{solute}}{\epsilon_{shell}}\right) V_{\Omega_{shell}}^{\Gamma_{diel}}\left(\frac{\partial\phi_{solute}}{\partial\mathbf{n}}\right) + V_{\Omega_{shell}}^{\Gamma_{diel}}\left(\frac{\sigma_{diel}}{\epsilon_{shell}}\right) \nonumber\\ 
-&\left(1-\frac{\epsilon_{solv}}{\epsilon_{shell}}\right) V_{\Omega_{shell}}^{\Gamma_{shell}}\left(\frac{\partial\phi_{solv}}{\partial\mathbf{n}}\right) + V_{\Omega_{solute}}^{\Gamma_{shell}}\left(\frac{\sigma_{shell}}{\epsilon_{shell}}\right).
\end{align}
In our model the potential is constant throughout the solute and the solvent, making the derivatives $\partial\phi_{solute}/\partial\mathbf{n}$ and $\partial\phi_{solv}/\partial\mathbf{n}$ zero everywhere.
Then, the first and third term in the right hand side of Equation \eqref{eq:phi_stern}, which are related to the jump in dielectric constant, disappear, and the potential is completely determined by the surface charge densities only, yielding
\begin{equation}\label{eq:surface_charge}
\phi(\mathbf{r}) = \oint_{\Gamma_{diel}} \frac{\sigma_{diel}(\mathbf{r}')}{4\pi \epsilon_{shell}|\mathbf{r}- \mathbf{r}'|} {d}\mathbf{r}' + \oint_{\Gamma_{shell}} \frac{\sigma_{shell}(\mathbf{r}')}{4\pi \epsilon_{shell}|\mathbf{r}- \mathbf{r}'|} {d}\mathbf{r}'.
\end{equation}
This is equivalent to the matrix representation in Equation (1) of the main text. 
It is interesting to note that this last step removes any influence of the permittivities of the solute (usually, $\epsilon_{solute}=1$) and solvent ($\epsilon_{solv}=80$).